\documentclass[10pt,aps,pre,twocolumn,superscriptaddress,showpacs]{revtex4-1}

\usepackage{amsfonts,amssymb,amsmath,latexsym,epsfig,wasysym} 
\usepackage[sort&compress]{natbib}
\usepackage{color}
\definecolor{bluecolor}{rgb}{0,0.,1.}

\definecolor{redcolor}{rgb}{.7,0.,0.}

\usepackage{graphicx}
\usepackage{dcolumn}
\usepackage{bm}
\usepackage{hyperref}

\definecolor{bluecolor}{rgb}{0,0.,1.}

\definecolor{redcolor}{rgb}{.7,0.,0.}

\begin{document}


\title{Using text analysis to quantify the similarity and evolution of scientific disciplines}

\author{La\'{e}rcio Dias}
\affiliation{Max Planck Institute for the Physics of Complex Systems, D-01187 Dresden, Germany}

\author{Martin Gerlach}
\affiliation{Max Planck Institute for the Physics of Complex Systems, D-01187 Dresden, Germany}
\affiliation{Department of Chemical and Biological Engineering, Northwestern University, Evanston, Illinois 60208, USA}

\author{Joachim Scharloth}
\affiliation{TU Dresden, Department of German, Applied Linguistics, D-01062 Dresden, Germany}

\author{Eduardo G.\ Altmann}
\affiliation{Max Planck Institute for the Physics of Complex Systems, D-01187 Dresden, Germany}
\affiliation{School of Mathematics and Statistics, University of Sydney, 2006 NSW, Australia}

\date{\today}


\begin{abstract}

We use an information-theoretic measure of linguistic similarity to investigate the
organization and evolution of scientific fields.  An analysis of almost 20M papers
from the past three decades reveals that the linguistic similarity is related
but different from experts and citation-based classifications, leading to an improved
view on the organization of science. A temporal analysis of the similarity of fields shows
that some fields (e.g., computer science) are becoming increasingly central, but that on average
the similarity between pairs has not changed in the last decades. This suggests that
tendencies of convergence (e.g., multi-disciplinarity) and divergence  (e.g.,
specialization) of disciplines are in balance.
\end{abstract}

\maketitle


\section{Introduction}
\label{sec:introduction}

The digitization of scientific production opens new possibilities for quantitative studies on scientometrics and science of science~\cite{evans.2011}, bringing new insights into questions such as how knowledge is organized (maps of science) \cite{boerner.2003,Shiffrin2004,boyack.2005,Rosvall2008,glaser.2017}, how impact evolves over time (bibliometrics) \cite{Wang2013,moreira.2015}, or how to measure the degree of interdisciplinarity \cite{lariviere.book2014,Noorden2015}. 
At the heart of these questions lies the problems of identifying scientific fields and how they relate to each other.
The difficulty of these problems, and the inadequacy of a purely essentialist approach, was clear to K. R. Popper already in the 1950's~\cite{popper.1952}:
{\it ``The belief that there is such a thing as physics, or biology, or archaeology, and that these 'studies' or 'disciplines' are distinguishable by the subject matter which they investigate, appears to me to be a residue from the time when one believed that a theory had to proceed from a definition of its own subject matter. But subject matter, or kinds of things, do not, I hold, constitute a basis for distinguishing disciplines.''}~\cite{popper.1952}.
Instead, he argued that disciplines have a cognitive and a social dimension~\cite{balsiger.book2005}, i.e. they
{\it ``are distinguished partly for historical reasons and reasons of administrative convenience (such as the organization of teaching and of appointments), and partly because the theories which we construct to solve our problems have a tendency to grow into unified systems.''}~\cite{popper.1952}.

On the one hand, the social dimension of scientific fields can be defined in terms of different institutions establishing stable recurring patterns of behavior~\cite{guntau.book1991}: 
producing and reproducing institutions such as research institutes and universities, communicative institutions such as scientific societies, journals or conferences, collecting institutions (journals, libraries), as well as directing institutions (ministries, scientific advisory boards), etc. All these institutions contribute to the formation, stabilization, and reproduction of a discipline as well as its distinction from others. 
On the other hand, the cognitive dimension has been specified 
in Ref.~\cite{guntau.book1991} as a number of fundamental invariants in the procedural knowledge, which lead to the categorical construction of scientific knowledge. 
If this process causes a change in the cognitive realm for an object of knowledge, it constitutes a certain discipline. 

The brief discussion above is sufficient to show that both the definition and relation between scientific fields depend on multiple dimensions (e.g., essentialist, social, and cognitive). Traditional (expert) classifications are mostly motivated by the {\it''subject matters''} under investigation and can be associated to an essentialist view. The empirical analysis of citation networks, an approach with a long tradition in scientometry~\cite{Garfield1964,DeSollaPrice1965}, can be regarded as capturing the social dimension (i.e. collecting institutions in the form of journals).
While citations offer valuable insights into the structure and dynamics of science, they thus reflect only one particular dimension of the relationship between publications (or scientists) largely ignoring the actual content of the scientific articles. 
In contrast, the cognitive dimension can be operationalized with the help of linguistic features (e.g., keywords as indicators for conceptual imprints of disciplines).
The increasing availability of full text of scientific articles (e.g. of Open Access journals) provides new opportunities to study the latter aspect in the form of written language.
Examples include i) the tracking of the spread of individual words (memes)~\cite{kuhn.2014} or ideas~\cite{chavalarias.2013}, ii) quantifying differences in the scientific discourse between subdomains in biomedical literature~\cite{lippincott.2011} or ``hard'' and ``soft'' science~\cite{evans.2016a}, or iii) efforts to combine citation and textual information~\cite{braam.1991,boerner.2003,vilhena.2014,silva.2016,sienkiewicz.2016}.

In this work we advance the idea that the organization and evolution of science should be studied through different, complementary, dimensions. 
We add a new methodology that provides a meaningful, language-based, organization of scientific disciplines based on written text, we study how it compares to classifications obtained from experts as well as citations, and we study the temporal evolution in the relation between different scientific disciplines.
More specifically, we introduce an unsupervised methodology to analyze the text of scientific articles.
Our methodology is based on an information-theoretic dissimilarity measure we proposed recently~\cite{Gerlach2016} (more technically, it is a generalized and normalized Jensen-Shannon divergence between two corpora). The main advantage of this measure is that it has an absolute meaning (i.e., it is not based on relative comparisons) and it is statistically more robust than traditional approaches~\cite{Gerlach2016,Altmann2017}, e.g. with respect to the detection of spurious trends due to rare words and increasing corpus sizes.
We measure the similarity between scientific fields based on $\approx 10^7$ abstracts from the last 3 decades (Web of Science database).  
Comparing our language analysis to a citation analysis and an experts classification, we find that the language and citation are more similar to each other but the language is even more distinct from the experts than the citation analysis. 
Following the relation between scientific fields over time, our language analysis reveals the scientific fields that are becoming  more central in science. 
However, overall (averaged over all pairs of disciplines) we find that the similarity between the language of different fields is not increasing.

\section{Dissimilarity measures of scientific fields}
\label{sec:dissimilarities}

 We are interested in the general problem~\cite{boerner.2003,boyack.2005} of quantifying the relationship between two
scientific  fields $i,j$ through the computation of dissimilarity measures
$D(i,j)$, i.e., a quantification of how different $i$ and $j$ are.
Dissimilarity measures are symmetric $D(i,j)=D(j,i)$, non-negative $D(i,j)\ge 0$, and $D(i,i)=0$~\cite{webb.2002}. 
Each scientific field is defined by (at least hundreds of) papers classified by Web of Science as belonging to the same category (see Methods Sec.~\ref{sec.method.data} for details on the data).
We consider dissimilarities computed based on the following three different information.

\subsection{Experts} 

The classification of disciplines by their relationship is as old as science itself. 
The most used structure is a strict hierarchical tree, as seen in the traditional departmental division of Universities. 
The collection of papers used here, provided by ISI Web Of Science~\cite{webOfScience}, provides a classification of papers according to the OECD classification of fields of science and technology~\cite{bibliographicWoSClassification}. 
This scheme is a hierarchical tree with scientific fields defined at 3 levels (domains, disciplines, and specialties).  
For instance, {\it Applied Mathematics} (a specialty) is part of {\it Mathematics} (a discipline) which is part of {\it Natural Sciences} (a domain).
The natural dissimilarity measure $D_{\text{exp}}(i,j)$ between two fields in this structure is the number of links needed to reach a common ancestor of $i$ and $j$. 
For instance, considering $i,j$ at the specialty level, $D_{\text{exp}}$ can assume three different values: $D_{\text{exp}}=1$ for specialties belonging to the same discipline (e.g., {\it Applied Mathematics} and {\it Statistics \& Probability}),
$D_{\text{exp}}=2$ for specialties belonging to the same  domain (e..g, {\it Applied Mathematics} and {\it Condensed Matter Physics}), and $D_{\text{exp}}=3$ for the other pairs of specialties
(e.g., {\it Applied Mathematics} and {\it Linguistics}).
While researchers have pointed out potential issues with classification into categories of ISI Web Of Science~\cite{boyack.2005}, it offers the most extensively available classification and remains widely used to relate articles and journals to disciplines~\cite{porter.2009,lariviere.book2014}. 

\subsection{Citations} 
Another popular approach is to consider that fields $i$ and $j$ are more similar if there are citations from (to) papers in $i$ to (from) papers in $j$~\cite{Garfield1964,DeSollaPrice1965,boyack.2005}.
Here we consider a dissimilarity measure $D_{\text{cite}}(i,j)$ which decreases for every citation between papers in $i$ and $j$, increases with every citation from $i$  that is not to $j$ (and vice-versa), but that remains unchanged by the number of citations that do not involve neither $i$ nor $j$. 
These requirements are achieved using (for $i \ne j$) a symmetrized Jaccard-like dissimilarity~\cite{webb.2002,leydesdorff.2008} 
\begin{equation}\label{eq.Dcitations}
D_{\text{cite}}(i,j) = \frac{1}{2}
\left(\frac{C_{i,\bar{j}}+C_{\bar{i},j}}{c_{i,j}+C_{i,\bar{j}}+C_{\bar{i},j}} +\frac{C_{j,\bar{i}}+C_{\bar{j},i}}{c_{j,i}+C_{j,\bar{i}}+C_{\bar{j},i}}\right)
\end{equation}
where $c_{i,j}$ are the number of citations from $i$ to $j$, $C_{a,\bar{b}} = \sum_{t=1,t\neq b}^N
c_{a,t}$, and $C_{\bar{a},b} = \sum_{t=1,t\ne a}^N c_{t,b}$\footnote{Each of the two terms in Eq.~(\ref{eq.Dcitations}) can be interpreted as a directed Jaccard distance $i \rightarrow j$ ($j \rightarrow i$) in the sense that we divide the number of edges that are out-links of field $i$ ($j$) \textit{and} in-links of field $j$ ($i$) by the number of edges that are out-links of field $i$ ($j$) \textit{or} in-links of field $j$ ($i$).}.

\begin{figure*}[!tb]
\centering
\includegraphics[width=0.43\textwidth]{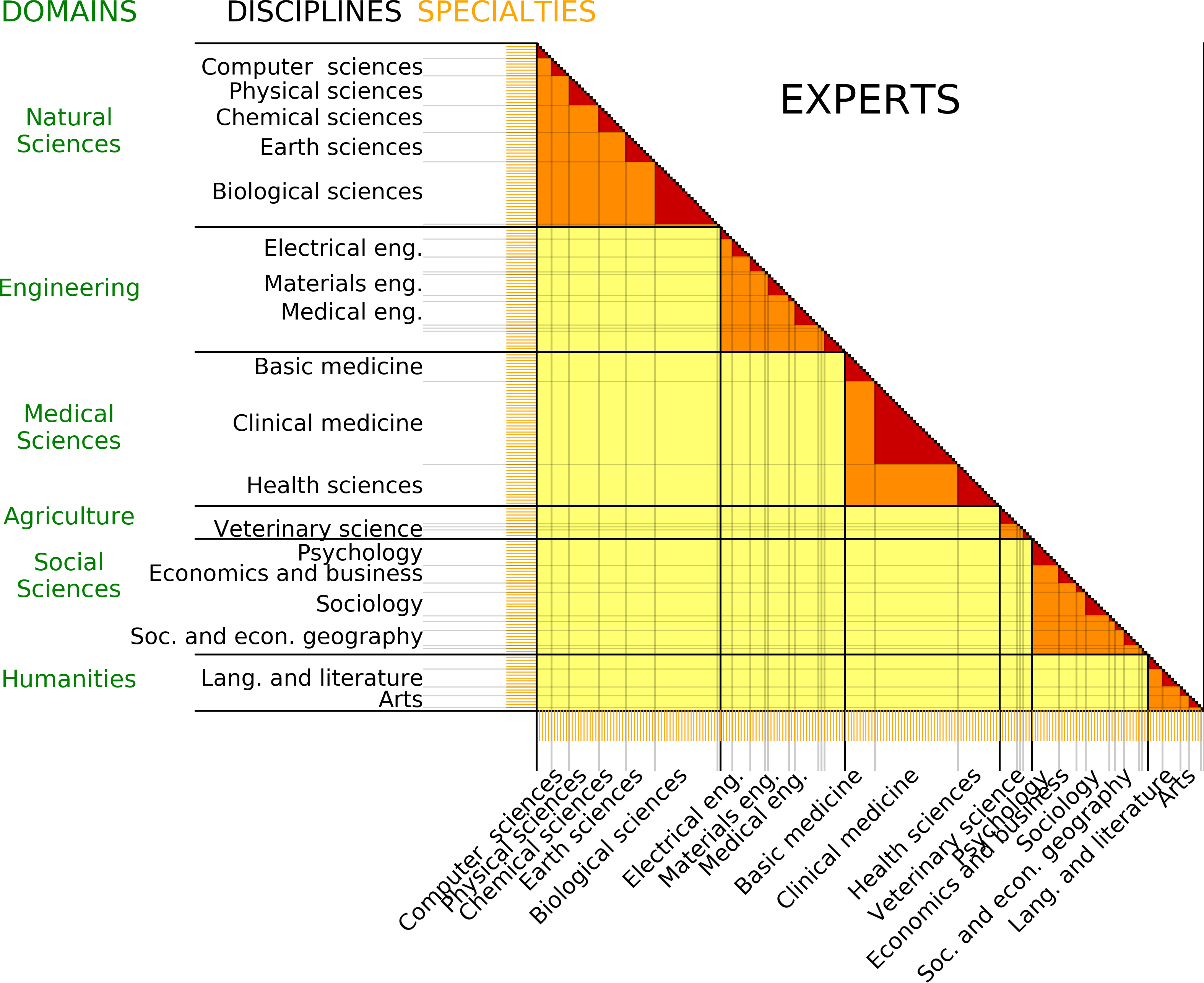}\hspace{-0.04\textwidth}\includegraphics[width=0.29\textwidth]{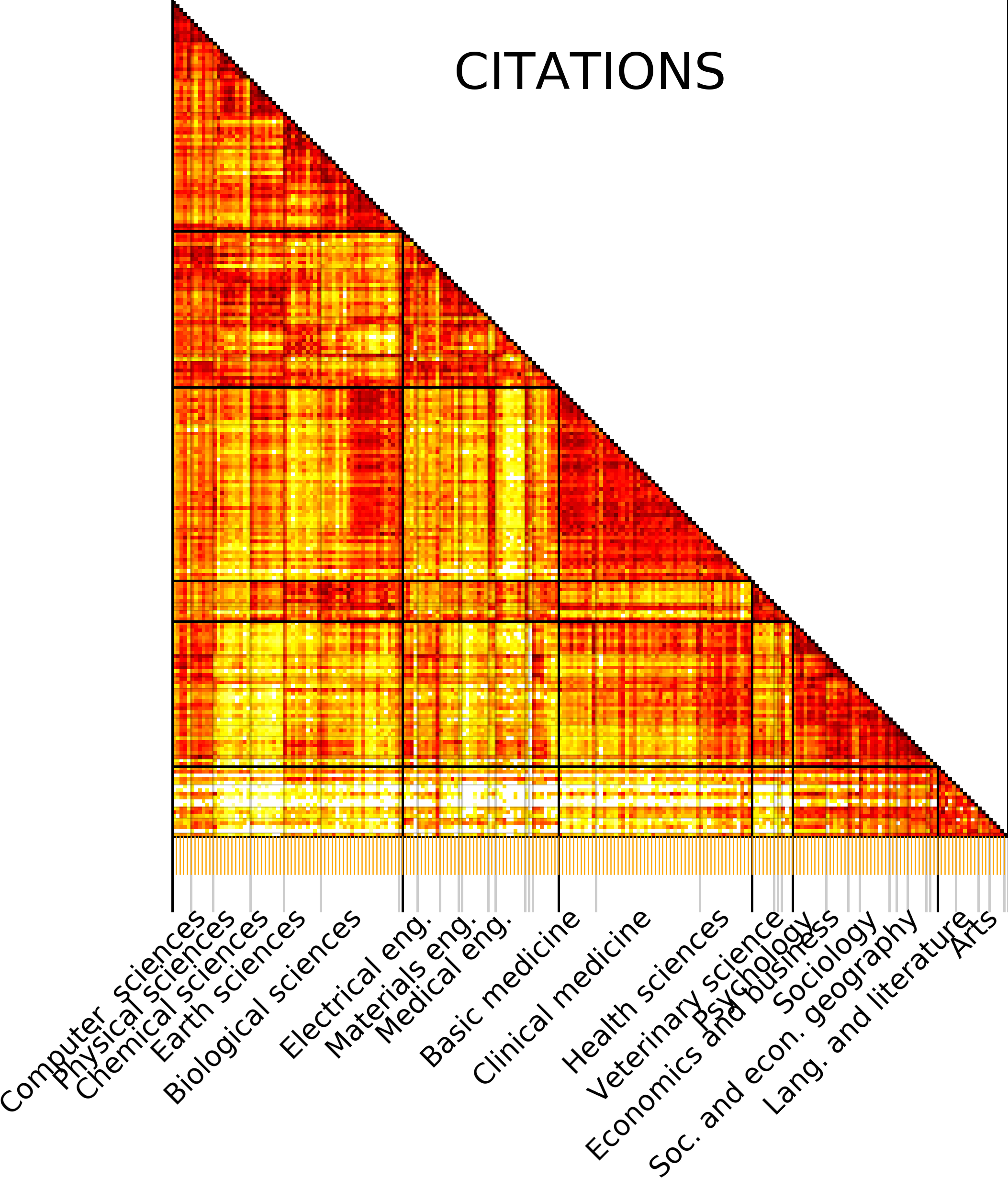}\hspace{-0.04\textwidth}\includegraphics[width=0.355\textwidth]{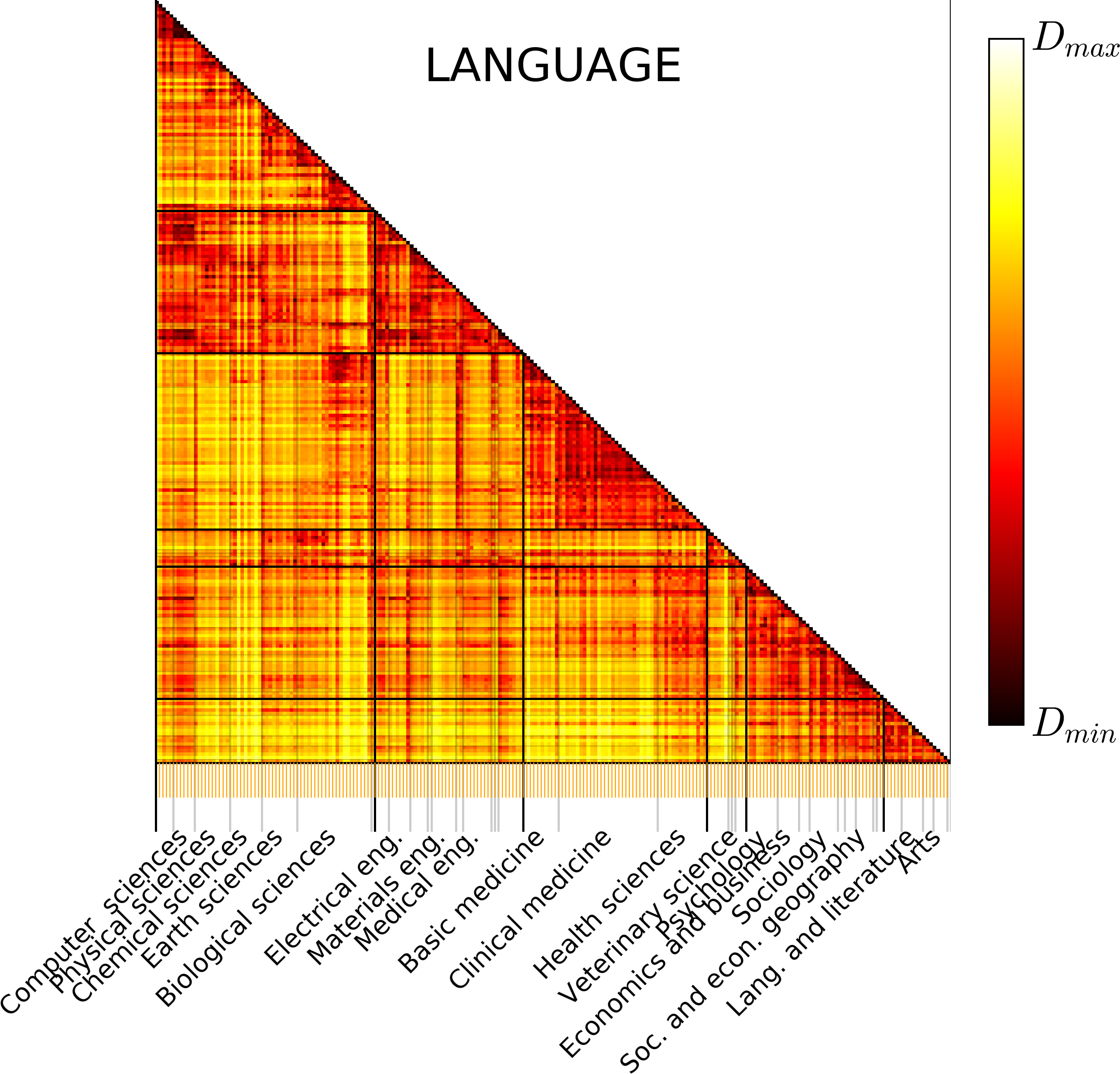}
\caption{Dissimilarity between specialties measured in three different dimensions: (a) $D_{\text{exp}}$ based on 
 experts classification ~\cite{bibliographicWoSClassification}, where $D_{min}=0$ and $D_{max}=4$; (b)  Citations dissimilarity $D_{\text{cite}}$~(\ref{eq.Dcitations}), where $D_{min}=0$ and $D_{max}=7.5$; (c) Language   dissimilarity $D_{\text{lang}}$~(\ref{eq.Dlanguage}), where $D_{min}=0$ and $D_{max}=1$. $N=225$  specialties of the OECD classification scheme are considered. Results based on $\approx 21M$ papers from $[1991,2014]$, see Sec.~\ref{sec.method.data} for details.}
\label{fig:distance_matrices}\label{fig.1}
\end{figure*}

\subsection{Language} 
We compare the language of fields $i$ and $j$ based on the frequency of words in each field using methods from Information Theory. 
Measuring the frequency $p(w)$ of word $w$, for each field $i$ we obtain a vector of frequencies $\mathbf{p}_i \equiv p_i(w)$ for $w=1, \ldots, V$, where $V$ is the size of the vocabulary (i.e. number of different words).
From this, following Ref.~\cite{Gerlach2016}, the dissimilarity between two fields $i$ and $j$ is
\begin{equation}\label{eq.Dlanguage}
D_{\text{lang}}(i,j) = \dfrac{2 H_{2}\left(\frac{\mathbf{p}_i+\mathbf{p}_j}{2}\right)
- H_{2}(\mathbf{p}_i) - H_2(\mathbf{p}_j)}{\frac{1}{2}\left(2- H_{2}(\mathbf{p}_i) - H_{2}(\mathbf{p}_j) \right)},
\end{equation}
where $ H_2 (\mathbf{p}_i) = 1 -  \sum_w p_i(w)^2 $ is the generalized entropy of order
$2$ and the denominator ensures normalization (i.e., $0 \le D_{\text{lang}}(i,j)\le 1$). 
In order to increase the discrimination power and to avoid statistical biases in our estimation, we removed a list of stop words 
and included only the $V=20,000$ most frequent words (see  Methods Sec.~\ref{sec.method.corpus.size} for a justification). The dissimilarity~(\ref{eq.Dlanguage}) corresponds to a generalized (and normalized) Jensen-Shannon divergence 
which yields statistically robust estimations in texts~\cite{Gerlach2016,Altmann2017} (for details and motivation, see  Methods Sec.~\ref{sec.method.sim.jsd}). 

The advantages of Eq.~(\ref{eq.Dlanguage}) are twofold.
On the one hand, it is well-founded in Information Theory and its statistical properties (in terms of systematic and statistical errors) are well understood~\cite{grosse.2002,Gerlach2016} distinguishing it from other heuristic approaches.  
On the other hand, it has convenient properties: i) $0 \le D_{\text{lang}}(i,j) \le 1$; ii) it depends only on the papers contained in fields $i$ and $j$; and iii) it does not require training corpora. 
As a result, the measured distance between two fields, $D_{\text{lang}}(i,j)$, has an absolute meaning.
This is in contrast to alternative similarity measures~\cite{boyack.2005,boerner.2003}, including machine-learning approaches (e.g., topic models~\cite{Landauer2004,Boyack2011}) based on (un-) supervised classification of documents into coherent subgroups.
Here, the main limitations stem from the fact that either i) the division into subgroups is typically based on statistically significant differences in the usage of words between the different subgroups independent of the actual effect size, or ii) the resulting distance between two fields depends on all other fields as well (e.g. the distance between 'Physics' and 'Chemistry' depends on whether one includes articles about 'Anthropology' in the classification). 

\section{Results}

We now present and interpret results obtained computing the three dissimilarity measures
($D_{\text{exp}}, D_{\text{cite}},$ and $D_{\text{lang}}$) reported above for scientific fields $i,j$ defined by papers published in different time intervals and categorized (by Web of Science) as belonging to the same specialty (e.g., Applied Mathematics), discipline, (e.g., Mathematics) or domain (e.g., Natural Sciences).

\subsection{Comparison of dissimilarity measures}

Figure~\ref{fig.1} shows the three $D(i,j)$ at the level of specialties $(i,j)$ for the complete time interval $1991-2014$.  
The concentration of low $D(i,j)$ close to the diagonal shows
that both the citations and language of scientific papers partially reflect the disciplinary classification done by the experts.
However, visual inspection already reveals that citations and our language analysis show relationships not present in the expert classification,
e.g., the low dissimilarity between Engineering and Natural Sciences (most clearly between Electrical Engineering and Physical Sciences) and between Agriculture and Biological Sciences.

We start by quantifying the relationship between the three different dissimilarity measures, i.e. ($D_{\text{exp}}, D_{\text{cite}},$ and $D_{\text{lang}}$), across all pairs of specialties $(i,j)$.
In Tab.~\ref{tab.rank}  we report the rank-correlation between the three measures, which we obtain from ranking for each dissimilarity the pairs of $(i,j)$ according to $D(i,j)$.
The choice of this non-parametric correlation is motivated by the fact that the range of the three measures differs dramatically (e.g. $D_{\text{exp}} \in \{0,1,2,3\}$ and $D_{\text{lang}} \in [0,1]$).
The positive statistically-significant correlation between all pairs of $D(i,j)$'s confirms the visual impression described above. 
The correlation between citations and language is higher than the correlation with the experts classification. Remarkably, language and citations show a very similar correlation with experts but language is systematically less correlated than citations ($p \text{-value}=1.8 \times 10^{-5}$ for Spearman-$\rho$ and  $p\text{-value}=2.2 \times 10^{-5}$ for Kendall-$\tau$~\footnote{Obtained from $10^3$ bootstrapping samples of each joint distribution $P(D_{\text{exp}},D_{\text{cite}})$ and $P(D_{\text{exp}},D_{\text{lang}})$, i.e. comparison of $10^6$ pairs of correlation values}). 
We conclude that the language dissimilarity $D_{\text{lang}}$ introduced here is able to retrieve the well-known relationships between disciplines in a similar extent that the (well-studied) citation analysis.

\begin{table}[hbt]
\renewcommand{\arraystretch}{1.5} 
\setlength{\tabcolsep}{3pt} 
\centering
\begin{tabular}{l | c c c}
\hline
\hline
Time & lang-cite & lang-exp & cite-exp \\
\hline
All, 1991-2014 & $0.57$ ($0.76$) & $0.32$ ($0.39$) & $0.34$ ($0.42$) \\
\hline
$1^{\text{st}}$ half, 1991-2002 & $0.60$ ($0.80$) & $0.34$ ($0.41$) & $0.37$ ($0.46$) \\
$2^{\text{nd}}$ half, x2003-2014 & $0.64$ ($0.84$) & $0.35$ ($0.43$) & $0.38$ ($0.47$) \\
\hline
\hline
\end{tabular}
\caption{Rank correlation  between the dissimilarities measures $D_x(i,j)$ obtained from different dimensions $x \in \{\text{exp (experts)}, \text{cite (citations)}, \text{lang (language)} \}$ computed over all specialty pairs $(i,j)$. 
All values are significantly different from zero (p-values  $< 10^{-5}$). 
The two values in each cell denote the Kendall-$\tau$ and Spearman-$\rho$ (in parenthesis).  
  Qualitatively equivalent results are obtained in three different time intervals (indicated in the left row).}
\label{tab.rank}
\end{table}

\begin{figure}[!htbp]
\centering
\includegraphics[width=0.47\textwidth]{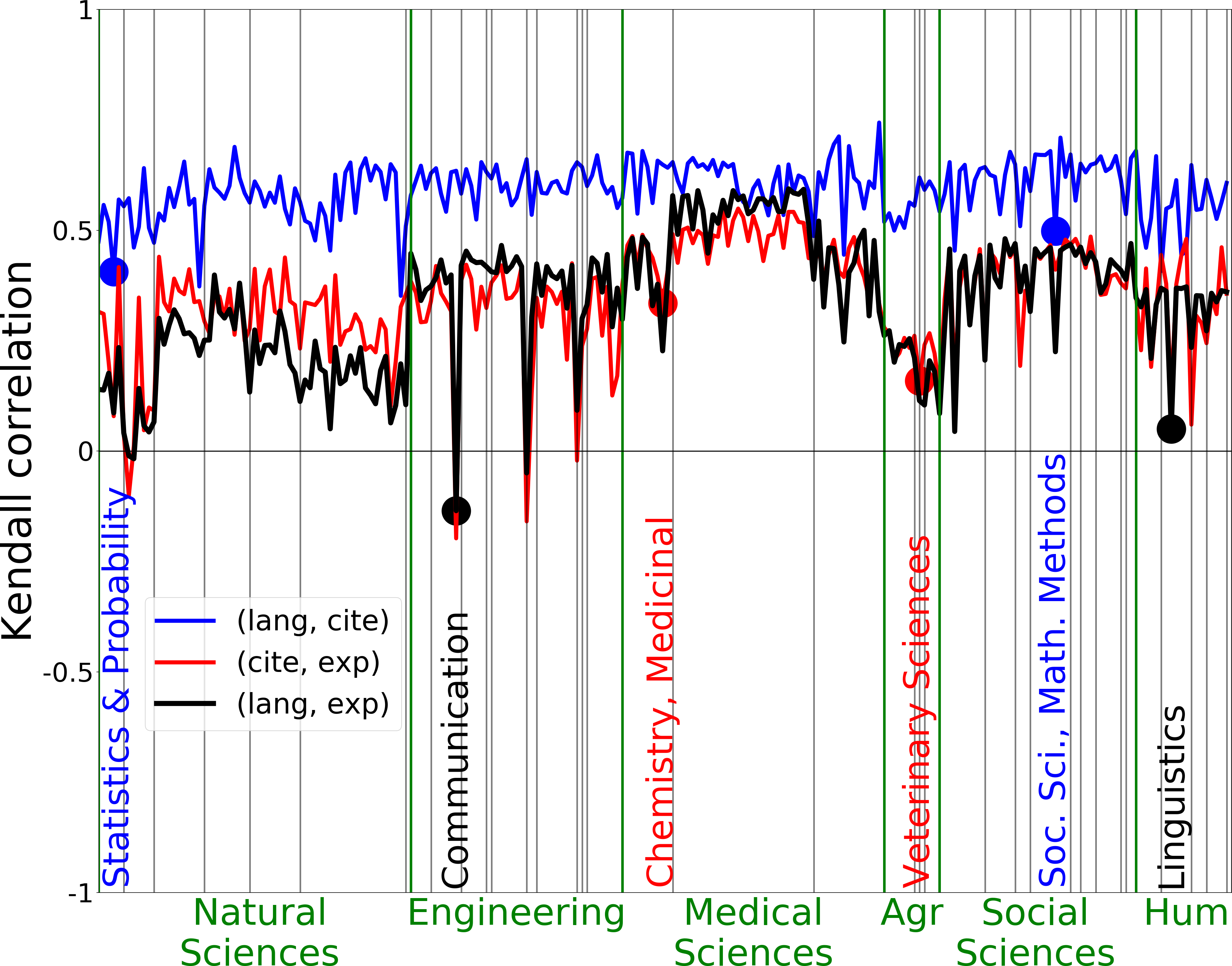}
\caption{Correlation between the different dissimilarity measures varies across fields. The Kendall correlation $\tau(x,y)$ (shown in the vertical axis) for two measures $x$ and $y$ is computed between $D_x(i,j)$ and $D_y(i,j)$ over all  specialties $j$  for a fixed specialty $i$ (shown in the horizontal-axis).
The three possible comparisons $(x,y)$ are indicated in the caption. Six specialties (one from each domain) with low correlation are highlighted.
  }
\label{fig:correlation_jsd2subfields_wos}\label{fig.kendall}
\end{figure}

We now explore how the relationship between the different dimensions depends on the different scientific fields. The results in Fig.~\ref{fig.kendall} confirm the conclusions of the aggregated analysis but shows further interesting features. First, the correlation in ($D_{\text{exp}},D_{\text{lang}}$) is smaller than ($D_{\text{exp}},D_{\text{cite}}$) mainly in the natural sciences. Second, while the correlation between citations and language remains largely constant, large fluctuations in the correlations between expert and citations (as well as expert and language) exist. This is seen both as the strong downward spikes and also in the  manifested dependence on disciplines and domains. 
The titles of the specialties at the low peaks already suggest that these are specialties with interdisciplinary connections. 
For instance, {\it Chemistry, Medicinal} is a specialty that (according to the experts classification) belongs to the discipline {\it Basic Medicine} and to the domain {\it Medical Science}. 
Therefore $D_{\text{exp}}=3$ between {\it Chemistry, Medicinal} and all specialties of the {\it Natural Sciences} (in particular, for all specialties from the discipline {\it  Chemical Sciences}).  
Instead, the dissimilarity measured by citations $D_{\text{cite}}$ and language $D_{\text{lang}}$ yield much smaller values revealing the proximity of {\it Chemistry, Medicinal} to the {\it Natural Sciences} thus explaining the low correlation in ($D_{\text{exp}},D_{\text{cite}}$) and in ($D_{\text{exp}},D_{\text{lang}}$).
The central role of the natural sciences in other disciplines explains also the other spikes: computing for a list of selected specialties $i=i^{\text{spikes}}$ the pairs $(i,j)$ which suffered the largest rank change we find that 9 from the the top 10 specialties which increased most in ranks (comparing $D_{\text{exp}}$ with $D_{\text{lang}}$) were from the domain {\it Natural Sciences}  ($5$ of them from the discipline {\it Chemical sciences}, including the top 2 specialties).

\begin{figure*}[!htbp]
\centering
\includegraphics[width=0.95\textwidth]{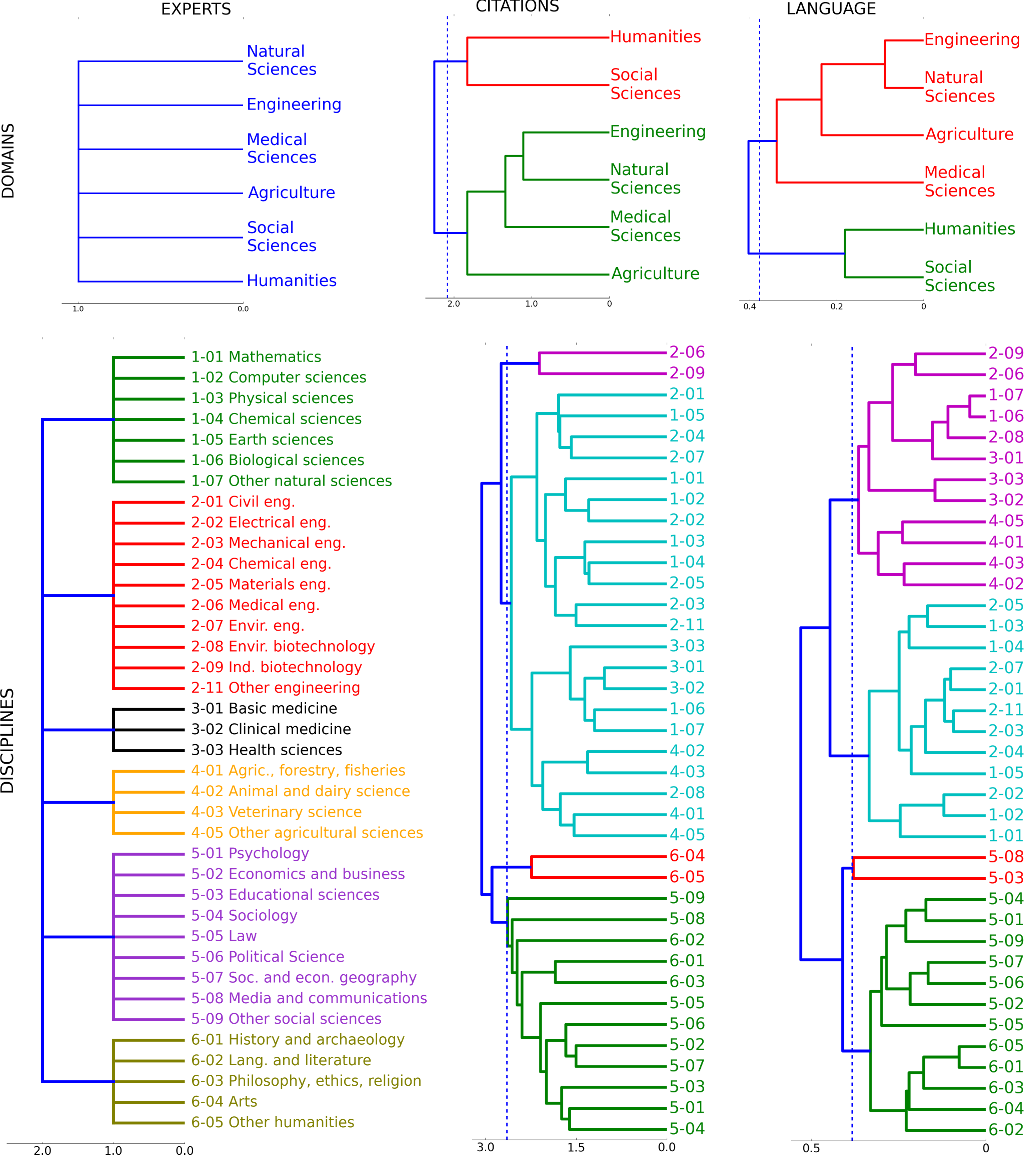}
\caption{Hierarchical clusterings at the level of domains (top row) and disciplines (bottom row). Results for citations (language) were obtained by agglomerative hierarchical clustering, applying the Group Average Method~\cite{Sokal1958} to  $D_{\text{cite}}(i,j)$ ($D_{\text{lang}}(i,j)$). The x-axis shows the clustering dissimilarity (i.e., the   dissimilarity of two clusterings that are merged). The colors reflect the clustering obtained at the dashed line, which corresponds to a clustering dissimilarity equals to the percentile 0.92 of the values of all cluster dissimilarities at each measure (citations/language).}
\label{fig:hierarchy}\label{fig_domains_hierarchy}\label{fig.3}
\end{figure*}

\subsection{Hierarchical Clustering}
\label{sec:subfieldsRelation}

A strict hierarchical classification of scientific fields is both aesthetically appealing and of practical use in bibliographical and document classification tasks. 
It also allows us to further highlight the differences in the relationship between scientific fields revealed by the different dissimilarity measures (in particular by $D_{\text{lang}}$).
While $D_{\text{exp}}$ is precisely based on one such hierarchical classifications, $D_{\text{cite}}$ and $D_{\text{lang}}$ are not. 
In Fig.~\ref{fig_domains_hierarchy} we show the hierarchical classifications induced  by
$D_{\text{cite}}$ and $D_{\text{lang}}$ through the computation of a simple clustering
method at the level of domains and disciplines. 

At the top level of the $6$ domains (top row in Fig.~\ref{fig_domains_hierarchy}), the clustering obtained from citations and from language are very similar. 
In particular, both identify {\it Engineering}-{\it Natural  Sciences} and  {\it Humanities}-{\it Social Science} as clusters that separate from the other domains in a similar fashion. 
The only difference is that, based on citations, {\it Agriculture} appears more isolated while based on language this happens for {\it Medical Science}. 
A more detailed picture of the differences between language and citation is revealed at the level of disciplines (bottom row in Fig.~\ref{fig_domains_hierarchy}).  
While at the first division, both citations and language create a cluster in which all disciplines of the domains {\it Humanities } and {\it Social Sciences} appear, further divisions show more subtle differences between the two dissimilarity measures.

Remarkably, the hierarchy obtained from language creates a cluster containing all and only Humanities disciplines.
In contrast, the hierarchy based on citations creates one clustering with three of the five {\it Humanities} disciplines ({\it Lang. and Literature}, {\it Arts}, and {\it Other Humanities}  while the two remaining ones ({\it History \& Archaeology} and {\it Philosophy, ethics, religion}) are clustered together in the middle of a cluster of disciplines in {\it Social Science}.
 Another interesting difference between the clusterings is revealed looking at 3 disciplines of the domain {\it  Medicine}: In the analysis based on Citations the minimum cluster that includes
the three disciplines includes {\it  Biological sciences} and {\it Other natural sciences},  while in the  language analysis this cluster includes additionally three related {\it Engineering} disciplines  ({\it Medical eng.}, {\it Ind. biotechnology}, and {\it Envir. biotechnology}).  

Probably the most remarkable feature of the clustering obtained by, both, citations and language is that it repeatedly clusters together related
disciplines from {\it Natural Sciences} with disciplines from {\it Engineering} and {\it Medicine} (e.g., {\it Chemical Sciences} and {\it Materials Science}). 
This clustering, not present in the experts classification, suggests that the distinction between fundamental and applied sciences present in the expert classification has no strong effect on citations and the language of the publications. Instead, in this specific case, the citation and language analysis seem to be capturing a connection between ``subject matters'' that was necessarily absent from the strict hierarchical expert classification.


\begin{figure}[!h]
\centering
\includegraphics[width=\columnwidth]{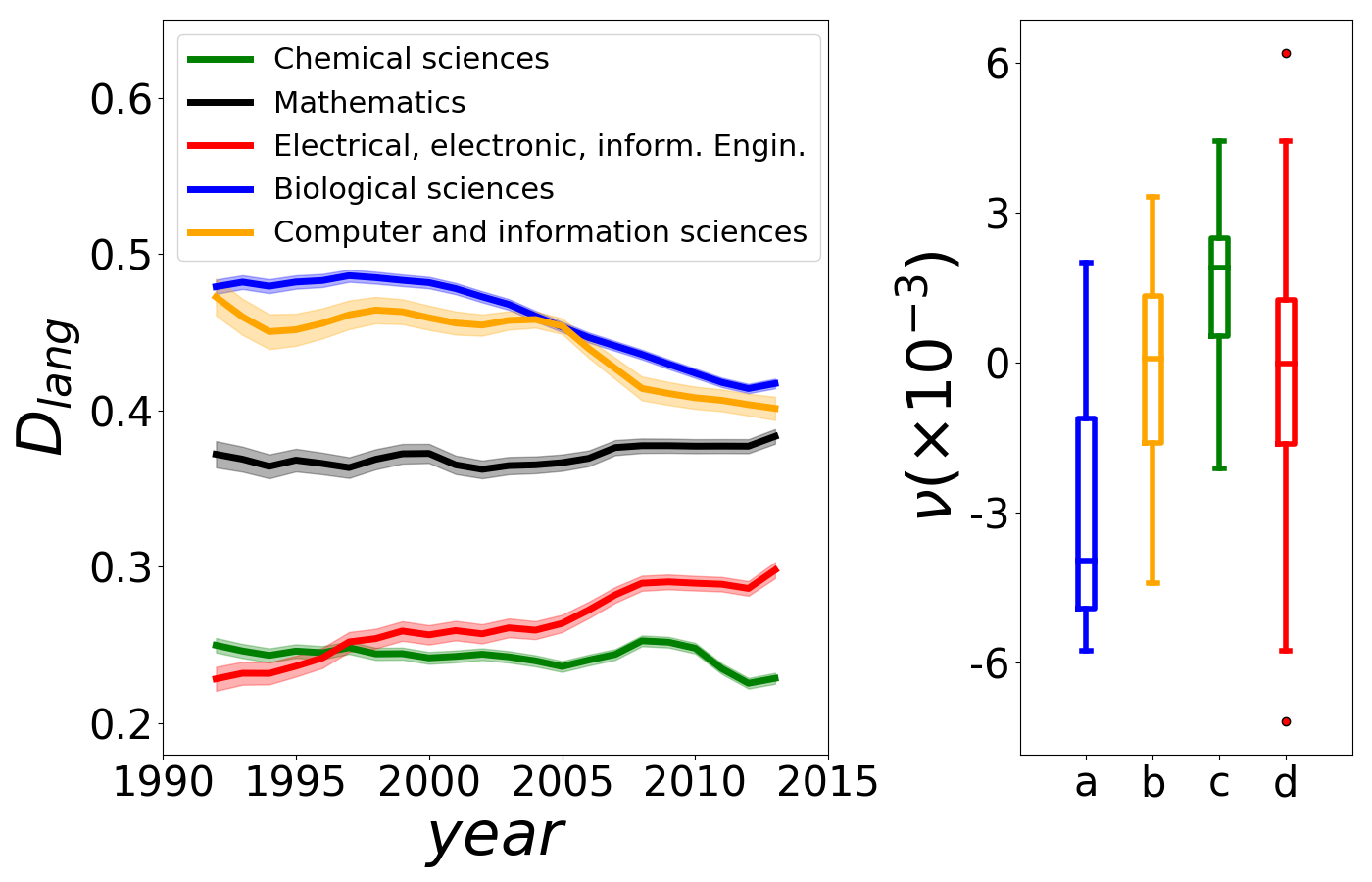}
\caption{Evolution of the similarity between disciplines in the last three decades. Left panel: distance $D_{\text{lang}}(i,j)$ between Physical Sciences ($i$) and other five selected disciplines ($j$, three-year moving averages). Right panel: total variation $\nu$ -- defined in Eq.~(\ref{eq.nu}) -- of the distance for pairs of disciplines with histories longer than $12$ years. Each boxplot corresponds to the distribution of $\nu$ for pairs of disciplines where we fixed one of the disciplines. At position (a) we fixed {\it Computer and information sciences}, at (b) {\it Chemical sciences}, at (c) {\it Psychology}, and at (d) we used all pairs of disciplines. 
}
\label{fig:evolution_jsd2_fields}\label{fig.4}
\end{figure}

\subsection{Temporal evolution}
\label{sec:fields_evolution}

While in the previous sections we looked at a static snapshot of the relation between disciplines, here we are interested in how the linguistic relationship $D_{\text{lang}}(i,j)$ between pairs $(i,j)$ of disciplines evolved over the last three decades~\footnote{We work at the level of disciplines because most specialties fail to have enough publications in a single year.}.
In Figure~\ref{fig:evolution_jsd2_fields} we show the temporal evolution for five out of $703$ pairs $(i,j)$, with focus on the discipline {\it Physical Sciences}, illustrating different types of dynamic patterns.
On the one hand, the dissimilarity to {\it Chemical Sciences} (its most similar discipline) and {\it Mathematics} stay roughly constant over time.
On the other hand, we also observe systematic trends of disciplines becoming more or less similar over time.
While the proximity to {\it Biological Sciences} and {\it Computer and information Science} has steadily increased (decreased dissimilarity $D_{\text{lang}}(i,j)$) after the year $2000$, the opposite trend is seen for  {\it Electrical, electronical, and information   Engineering}.
These observations are consistent with the increasing number of biological and computational-related publications in Physics, and with a departure from the historical connections to Engineering.

The observations reported above raise the question whether scientific disciplines are showing an overall tendency to become more similar to each other.
In a more general context, this amounts to the question whether the purported increase in {\it interdisciplinarity} leads to a larger overlap in the language used by different disciplines. 
We address this question by computing, for each pair of disciplines, the mean yearly variation
\begin{eqnarray}\label{eq.nu}
\nu(i,j) &= \dfrac{1}{\Delta t} \sum_{t \in \Delta t}  D_{\text{lang}}^{(t)}(i,j)- D_{\text{lang}}^{(t-1)}(i,j) \\
&=   \dfrac{1}{\Delta t} \left(D_{\text{lang}}^{(t_f)}(i,j) - D_{\text{lang}}^{(t_0-1)}(i,j)\right),
\end{eqnarray}
where the time interval $\Delta t\equiv t_f-t_0$ was usually from $t_0=1991$ to $t_f=2014$. 
The distribution of values of $\nu$ for all disciplines pairs $(i,j)$ is shown at the (rightmost) box plot in the right panel of Fig.~\ref{fig:evolution_jsd2_fields}.
We see that there are both positive and negative variations, consistent with our qualitative observations in the example of {\it Physical Sciences} in left panel of the Fig.~\ref{fig:evolution_jsd2_fields}.
However, the average variation $\langle \nu \rangle \approx -0.00025$ over all pairs of disciplines $(i,j)$ is not distinguishable from zero (the null hypothesis of $\langle \nu \rangle=0$ has a p-value=$0.07$ in the T-test for the mean of one sample and a p-value = $0.21$ in the  non-parametric Wilcoxon test), i.e.  the typical dissimilarity remains unchanged.  
This result suggests that, while there are systematic trends for individual pairs of disciplines, on average there is no significant increase or decrease in the interdisciplinarity for the science as a whole in the last 3 decades as measured by the language. 

On a more fine-grained level, however, we observe systematic trends that suggest that individual disciplines tend to become more (less) central.
For this, we focus on the discipline pairs $(i,j)$ which experienced the most extreme variation in the last decade (one standard deviation away from $\langle \nu \rangle$).
These pairs have typically $\vert \nu \vert \gtrapprox 0.003$ meaning that their (normalized) dissimilarity changes roughly $3\%$ in a decade. 
The three disciplines that are most frequently seen in the left tail ($\nu < 0$) are: {\it 1-02 Computer and information sciences}, {\it 2-08 Environmental biotechnology}, and 3-01 {\it Basic medicine}. 
The language of these disciplines became significantly more similar to the language of other disciplines in the last 3 decades, suggesting that these disciplines became more central.
In contrast, the three disciplines that experienced most strongly the opposite effect (most frequently seen in the right tail, $\nu > 0$) are: 5-01 {\it Psychology}, 2-05 {\it Materials  engineering}, and  2-02 {\it Electrical engineering, electronic engineering, information  engineering}.


\section{Discussion}
\label{sec:conclusions}

We investigated the similarity between scientific fields from different perspectives: an
expert classification, a citation analysis, and a newly proposed measure of linguistic
similarity. We found that these different dimensions are related yet different,
yielding thus new insights on the relationship between disciplines, their hierarchical
organization, and their temporal evolution. 

Our first main finding is that the language and citation relationships between disciplines are similar and substantially different from the expert classification. This is consistent with the motivation exposed in our introduction which associated the expert classification to the (largely idealized) essentialist view of scientific disciplines, while the citation (social) and language (cognitive) were closer to dimensions that play a more important role in the relationship between fields. Interestingly, our results indicate that the language-relation of fields is more distinct from the expert classification than the citation-relation is, specially in the natural sciences.

Our second main finding is that in the last 30 years the language of different scientific fields remain, on average, at the same distance from all other fields. 
While individual disciplines show clear trends of increasing (or decreasing) centrality, this suggests that, overall, diverging tendencies in science (e.g., specialization) are in balance with converging tendencies (e.g., multidisciplinarism). 
This is a remarkable quantitative finding because of the substantial changes observed in this period.

The latter result demonstrates that our textual measure is of practical relevance for the study of interdisciplinarity.
In recent years, interdisciplinary research achieved a central position~\citep{Noorden2015}
due to its broader relation to the concept of diversity~\cite{stirling.2007} and its effect on impact~\cite{uzzi.2013,wang.2015} and performance of teams~\cite{lungeanu.2014} as well as its implications for policy making, e.g. in terms of funding~\cite{academy.book2004}.
Is it just a fashion or science is really getting more and more interdisciplinary? 
A usual way to assess interdisciplinarity is based on citation networks using heuristic approaches~\cite{porter.2009,wagner.2011,lariviere.book2014} or methods from complex networks~\cite{pan.2012,sayama.2012,sinatra.2015,omodei.2016}. In line with the arguments exposed in the introduction, interdisciplinarity can be viewed through different dimensions and the cognitive dimension would be best measured using textual data. However, there are only very few works~\cite{bache.2013,nichols.2014,evans.2016} relating textual measures with interdisciplinarity, despite the increasing availability of the text of scientific articles. In this view, the significance of our approach is that it provides a measure of interdisciplinarity based on how much the usage of words in different disciplines overlap.

Finally, we hope our results and methodology will stimulate a multiple-dimensional approach in other problems related to the study of sciences, profiting from the modern availability of large (textual) databases of scientific publications that allow us to go beyond traditional bibliometric analysis~\cite{evans.2011,lariviere.book2014}. 
These include, but are not limited to, the formulation of more meaningful bibliometric indicators~\cite{mann.2006},
the identification and prediction of influential papers and disciplines~\cite{gerrish.2010,foulds.2013,whalen.2015}, 
or the inclusion of textual information in recommending related scientific papers~\cite{achakulvisut.2016}.

\section{Materials and Methods}

\subsection{Data and grouping of corpora}
\label{sec.method.data}
We use the Web of Science database~\cite{webOfScience} and explore the following information available for individual articles: citations, title, abstract, and the classification in one scientific specialty (per OECD classification \cite{bibliographicWoSClassification}). We use all papers published between 1991 and 2014 because the number of articles with text in the abstract is substantial only after 1991 and because at the time we started our analysis 2014 was the last complete year available to us. The text of an article was built concatenation its title and abstract. The corpus representing a specialty in a given year is obtained from the concatenation of the text of all articles for that specialty in that year. The corpus for one discipline (or domain) concatenates all articles in all specialties belonging to that discipline (or domain).

Our analysis is based on $19,589,166$ articles for each the textual and classification information were available ($92\%$ of all articles indexed in Web of Science between 1991-2014). In our analysis we considered only citations from and to the papers in our list because only for these papers we had a reliable classification of specialties. These citations corresponded to roughly half of the $\approx 625$M citations associated with these papers. 

\subsection{Data processing}
\label{sec.method.filter}
For each article in our database we performed the following steps to process the textual information:

\begin{enumerate}
\item The copyright information contained in the abstract was removed.
\item Title and abstract were concatenated.
\item The text was converted to lowercase.
\item Contractions were replaced by their non-contracted form. 
\item The text was tokenized, and the nouns and verbs were lemmatized using the Natural Language Toolkit~\cite{nltk}.
\item Symbols (except hyphen, to avoid remove significant compound modifiers) inside tokens were replaced by white space, therefore generating  two or more distinct tokens.
\item Tokens composed by numbers or single letter were removed.
\item Tokens belonging to a preset stop-word list were discarded. 
\end{enumerate}

\subsection{Minimum corpus size}
\label{sec.method.corpus.size}

We computed $D_{\text{lang}}$ using only the $20,000$ most frequent word types, disregarding the scientific fields for which there was not enough data to achieve this cut-off. This choice is motivated by the slow convergence of entropy estimations (and thus $D_{\text{lang}}$)~\cite{Gerlach2016}. By choosing a fixed number of word types we reduce the effect of the remaining bias (in the estimation of $D_{\text{lang}}$) on our comparative analysis of textual dissimilarity between pairs of fields. This happens because the residual bias acts as an off-set in all cases (when a fixed cut-off is chosen) instead of affecting differently each case (as obtained if the maximum amount of data is used in each case). The bias decays with the number of word types used because the more frequent types are responsible for almost all the dissimilarity, specially for $\alpha=2$~\cite{Altmann2017}. Using $10,000$ types as a cut-off, we estimated the textual dissimilarity relative standard deviation, computed over multiple samples of the same scientific field, to be $\hat{\sigma}(D_{lang})/D_{lang} \approx 1\%$. Our cut-off of $20,000$ types is a conservative choice to ensure that $\hat{\sigma}(D_{lang})/D_{lang} < 1\%$.

\subsection{Generalized Jensen-Shannon Divergence}
\label{sec.method.sim.jsd}
Given two texts (indexed by $p$ and $q$), we define the probability distributions over all words $w=1,\ldots,V$  as $\mathbf{p} = (p_{w})$ and $\mathbf{q} = (q_{w})$.
An Information-theoretic measure to quantify their similarity is the generalized Jensen-Shannon divergence
\begin{equation}
D_{\alpha}(\mathbf{p}, \mathbf{q}) = H_{\alpha}\left(\frac{\mathbf{p}+\mathbf{q}}{2}\right) - \frac{1}{2}H_{\alpha}(\mathbf{p}) - \frac{1}{2}H_{\alpha}(\mathbf{q}),
\label{eq:jsd}
\end{equation}
based on the generalized entropy of order $\alpha$ ($\in \mathbb{R}$), where 
\begin{equation}
H_\alpha(\mathbf{p}) = \frac{1}{1-\alpha} \left( \sum_w p_w^\alpha - 1 \right).
\label{eq:Ha}
\end{equation}

Here, we consider a normalized similarity~\cite{Gerlach2016}
\begin{equation}
\tilde{D}_{\alpha}(\mathbf{p}, \mathbf{q}) = \frac{D_{\alpha}(\mathbf{p}, \mathbf{q})}{D_{\alpha}^{\max}(\mathbf{p}, \mathbf{q})}
\label{eq:Da_norm}
\end{equation}
such that $\tilde{D}_{\alpha} \in [0,1]$ where
$D_{\alpha}^{\max}(\mathbf{p}, \mathbf{q}) = \frac{2^{1-\alpha} - 1}{2} \left( H_{\alpha}(\mathbf{p}) + H_{\alpha}(\mathbf{q}) + \frac{2}{1-\alpha}\right)$
is the maximum possible $D_\alpha$ between $\mathbf{p}$ and $\mathbf{q}$ assuming that the the set of symbols in each distribution (i.e., the support of $\mathbf{p}$ and $\mathbf{q}$) are disjoint.

Note that for $\alpha = 1$, Eq.~\eqref{eq:Ha} yields the Shannon-entropy~\citep{cover2006}, i.e. $H_{\alpha=1}(\mathbf{p}) = -\sum_w p_w\log p_w$, and $D_{\alpha=1}$ is the well-known Jensen-Shannon divergence~\cite{lin1991}.
Ref.~\cite{Gerlach2016} shows that $\alpha=2$ provides the most robust statistical measure of similarity of texts.

\begin{acknowledgments}

L.D. received financial support from CNPq/Brazil through the program ``Science without
Borders''. We thank M. Palzenberger and the Max Planck Digital Library for providing access to the data, M. de Domenico for insightful discussions, and S. Haan and the Centre for Translational Data Science (University of Sydney) for helping with Figs.~\ref{fig.1} and~\ref{fig.3}.

\end{acknowledgments}


\begin{thebibliography}{60}

\bibitem{evans.2011}
J.~A. Evans, J.~G. Foster, {\it Science\/} {\bf 331}, 721 (2011).

\bibitem{boerner.2003}
K.~B{\"o}rner, C.~Chen, K.~W. Boyack, {\it Annual review of information science
  and technology\/} {\bf 37}, 179 (2003).

\bibitem{Shiffrin2004}
R.~M. Shiffrin, K.~Borner, {\it Proceedings of the National Academy of
  Sciences\/} {\bf 101}, 5183 (2004).

\bibitem{boyack.2005}
K.~W. Boyack, R.~Klavans, K.~B\"{o}rner, {\it Scientometrics\/} {\bf 64}, 351
  (2005).

\bibitem{Rosvall2008}
M.~Rosvall, C.~T. Bergstrom, {\it Proceedings of the National Academy of
  Sciences\/} {\bf 105}, 1118 (2008).

\bibitem{glaser.2017}
J.~Gl{\"{a}}ser, W.~Gl{\"{a}}nzel, A.~Scharnhorst, {\it Scientometrics\/} {\bf
  111}, 979 (2017).

\bibitem{Wang2013}
D.~Wang, C.~Song, A.-L. Barab{\'{a}}si, {\it Science\/} {\bf 342}, 127 (2013).

\bibitem{moreira.2015}
J.~A.~G. Moreira, X.~H.~T. Zeng, L.~A.~N. Amaral, {\it PLoS ONE\/} {\bf 10},
  e0143108 (2015).

\bibitem{lariviere.book2014}
V.~Larivi{\`{e}}re, Y.~Gingras, {\it Beyond Bibliometrics\/} (MIT Press, 2014).

\bibitem{Noorden2015}
R.~V. Noorden, {\it Nature\/} {\bf 525}, 306 (2015).

\bibitem{popper.1952}
K.~R. Popper, {\it The British Journal for the Philosophy of Science\/} {\bf
  3}, 124 (1952).

\bibitem{balsiger.book2005}
P.~W. Balsiger, {\it {Transdisziplinarit{\"{a}}t : systematisch-vergleichende
  Untersuchung disziplinen{\"{u}}bergreifender Wissenschaftspraxis}\/} (Fink,
  2005).

\bibitem{guntau.book1991}
M.~Guntau, H.~Laitko, {\it World Views and Scientific Discipline Formation\/},
  R.~W. Woodward, R.~S. Cohen, eds. (Springer Netherlands, 1991).

\bibitem{Garfield1964}
E.~Garfield, I.~H. Sher, R.~J. Torpie, {\it The use of citation data in writing
  the history of science\/} (Institute for Scientific Information,
  Philadelphia, 1964).

\bibitem{DeSollaPrice1965}
D.~J. de~Solla~Price, {\it Science\/} {\bf 149}, 510 (1965).

\bibitem{kuhn.2014}
T.~Kuhn, M.~Perc, D.~Helbing, {\it Physical Review X\/} {\bf 4}, 041036 (2014).

\bibitem{chavalarias.2013}
D.~Chavalarias, J.-P. Cointet, {\it {PLoS} {ONE}\/} {\bf 8}, e54847 (2013).

\bibitem{lippincott.2011}
T.~Lippincott, D.~{\'{O}}. S{\'{e}}aghdha, A.~Korhonen, {\it BMC
  bioinformatics\/} {\bf 12}, 212 (2011).

\bibitem{evans.2016a}
E.~Evans, C.~Gomez, D.~McFarland, {\it Sociological Science\/} {\bf 3}, 757
  (2016).

\bibitem{braam.1991}
R.~R. Braam, H.~F. Moed, A.~F.~J. van Raan, {\it Journal of the American
  Society for Information Science\/} {\bf 42}, 233 (1991).

\bibitem{vilhena.2014}
D.~Vilhena, {\it et~al.\/}, {\it Sociological Science\/} {\bf 1}, 221 (2014).

\bibitem{silva.2016}
F.~N. Silva, D.~R. Amancio, M.~Bardosova, L.~d.~F. Costa, O.~N. Oliveira, {\it
  Journal of Informetrics\/} {\bf 10}, 487 (2016).

\bibitem{sienkiewicz.2016}
J.~Sienkiewicz, E.~G. Altmann, {\it Royal Society Open Science\/} {\bf 3},
  160140 (2016).

\bibitem{Gerlach2016}
M.~Gerlach, F.~Font-Clos, E.~G. Altmann, {\it Physical Review X\/} {\bf 6}, 021009
  (2016).

\bibitem{Altmann2017}
E.~G. Altmann, L.~Dias, M.~Gerlach, {\it Journal of Statistical Mechanics:
  Theory and Experiment\/} {\bf 2017}, 014002 (2017).

\bibitem{webb.2002}
A.~Webb, {\it {Statistical Pattern Recognition}\/} (Wiley, 2002).

\bibitem{webOfScience}
\href{http://apps.webofknowledge.com}{Web of Science} is a product of
  \href{http://thomsonreuters.com/en.html}{Thomson Reuters}.

\bibitem{bibliographicWoSClassification}
Working Party of National Experts on Science and Technology, OECD (2006)
  available at \url{http://www.oecd.org/science/inno/38235147.pdf}.

\bibitem{porter.2009}
A.~L. Porter, I.~Rafols, {\it Scientometrics\/} {\bf 81}, 719 (2009).

\bibitem{leydesdorff.2008}
L.~Leydesdorff, {\it Journal of the American Society for Information Science
  and Technology\/} {\bf 59}, 77 (2008).

\bibitem{Note1}
Each of the two terms in Eq.~(\ref {eq.Dcitations}) can be interpreted as a
  directed Jaccard distance $i \rightarrow j$ ($j \rightarrow i$) in the sense
  that we divide the number of edges that are out-links of field $i$ ($j$)
  \protect \textit {and} in-links of field $j$ ($i$) by the number of edges
  that are out-links of field $i$ ($j$) \protect \textit {or} in-links of field
  $j$ ($i$).

\bibitem{grosse.2002}
I.~Grosse, {\it et~al.\/}, {\it Physical Review E\/} {\bf 65}, 041905 (2002).

\bibitem{Landauer2004}
T.~K. Landauer, D.~Laham, M.~Derr, {\it Proceedings of the National Academy of
  Sciences\/} {\bf 101 Suppl}, 5214 (2004).

\bibitem{Boyack2011}
K.~W. Boyack, {\it et~al.\/}, {\it PLoS One\/} {\bf 6}, e18029 (2011).

\bibitem{Note2}
Obtained from $10^3$ bootstrapping samples of each joint distribution
  $P(D_{\protect \text {exp}},D_{\protect \text {cite}})$ and $P(D_{\protect
  \text {exp}},D_{\protect \text {lang}})$, i.e. comparison of $10^6$ pairs of
  correlation values.

\bibitem{Sokal1958}
R.~Sokal, C.~Michener, {\it University of Kansas Science Bulletin\/} {\bf 38},
  1409 (1958).

\bibitem{Note3}
We work at the level of disciplines because most specialties fail to have
  enough publications in a single year.

\bibitem{stirling.2007}
A.~Stirling, {\it Journal of The Royal Society Interface\/} {\bf 4}, 707
  (2007).

\bibitem{uzzi.2013}
B.~Uzzi, S.~Mukherjee, M.~Stringer, B.~Jones, {\it Science (New York, N.Y.)\/}
  {\bf 342}, 468 (2013).

\bibitem{wang.2015}
J.~Wang, B.~Thijs, W.~Gl{\"{a}}nzel, {\it PLoS ONE\/} {\bf 10}, e0127298
  (2015).

\bibitem{lungeanu.2014}
A.~Lungeanu, Y.~Huang, N.~S. Contractor, {\it Journal of Informetrics\/} {\bf
  8}, 59 (2014).

\bibitem{academy.book2004}
{Committee on Facilitating Interdisciplinary Research; Committee on Science,
  Engineering}, P.~P.~I. of~Medicine;~Policy, G.~A. N.~A. of~Sciences; National
  Academy~of Engineering, {\it {Facilitating Interdisciplinary Research}\/}
  (National Academies Press, 2004).

\bibitem{wagner.2011}
C.~S. Wagner, {\it et~al.\/}, {\it Journal of Informetrics\/} {\bf 5}, 14
  (2011).

\bibitem{pan.2012}
R.~K. Pan, S.~Sinha, K.~Kaski, J.~Saram{\"{a}}ki, {\it Scientific Reports\/}
  {\bf 2}, 1 (2012).

\bibitem{sayama.2012}
H.~Sayama, J.~Akaishi, {\it PLoS ONE\/} {\bf 7}, e38747 (2012).

\bibitem{sinatra.2015}
R.~Sinatra, P.~Deville, M.~Szell, D.~Wang, A.-L. Barab{\'{a}}si, {\it Nature
  Physics\/} {\bf 11}, 791 (2015).

\bibitem{omodei.2016}
E.~Omodei, M.~D. Domenico, A.~Arenas, {\it Network Science\/} pp. 1--12 (2016).

\bibitem{bache.2013}
K.~Bache, D.~Newman, P.~Smyth, {\it Proceedings of the 19th ACM SIGKDD
  international conference on Knowledge discovery and data mining - KDD '13\/}
  (ACM Press, 2013).

\bibitem{nichols.2014}
L.~G. Nichols, {\it Scientometrics\/} {\bf 100}, 741 (2014).

\bibitem{evans.2016}
E.~D. Evans, {\it Socius: Sociological Research for a Dynamic World\/} {\bf 2}
  (2016).

\bibitem{mann.2006}
G.~S. Mann, D.~Mimno, A.~McCallum, {\it Proceedings of the 6th ACM/IEEE-CS
  joint conference on Digital libraries - JCDL '06\/} (ACM Press, 2006).

\bibitem{gerrish.2010}
S.~Gerrish, D.~M. Blei, {\it Proceedings of the 27th International Conference
  on Machine Learning (ICML-10), June 21-24, 2010, Haifa, Israel\/} (2010), pp.
  375--382.

\bibitem{foulds.2013}
J.~Foulds, P.~Smyth, {\it Proceedings of the 2013 Conference on Empirical
  Methods in Natural Language Processing\/} pp. 113--123 (2013).

\bibitem{whalen.2015}
R.~Whalen, Y.~Huang, A.~Sawant, B.~Uzzi, N.~Contractor, {\it Quantifying and
  Analysing Scholarly Communication on the Web (ASCW'15)\/} (2015).

\bibitem{achakulvisut.2016}
T.~Achakulvisut, D.~E. Acuna, T.~Ruangrong, K.~Kording, {\it PLoS ONE\/} {\bf
  11}, e0158423 (2016).


\bibitem{nltk}
Natural language toolkit, \url{http://www.nltk.org/}.

\bibitem{cover2006}
T.~M. Cover, J.~A. Thomas, {\it Elements of Information Theory\/}
  (Wiley-Interscience, 2006).

\bibitem{lin1991}
{J. Lin}, {\it IEEE Transactions on Information Theory\/} {\bf 37}, 145 (1991).

\end{thebibliography}

\end{document}